\documentclass[%
reprint, superscriptaddress,
showpacs,
nofootinbib,
 amsmath,amssymb,
 aps,
]{revtex4-1}

\usepackage[nodisplayskipstretch]{setspace}
\usepackage{cancel}
\usepackage{mathtools}
\allowdisplaybreaks
\usepackage{tabularx}

\usepackage{amssymb}
\usepackage{amsmath}
\usepackage{epsfig}
\usepackage[hidelinks]{hyperref}
\usepackage{breakurl}
\usepackage{bm}
\usepackage{color}
\usepackage{tikz}

\newcommand{\cO}{\mathcal{O}}
\newcommand{\cM}{\mathcal{M}}

\newcommand{\cE}{\mathcal{E}}
\newcommand{\ex}{\text{e}}
\newcommand{\nl}{\notag\\[1ex]}
\makeatletter
\newcommand{\vast}{\bBigg@{3}}
\newcommand{\Vast}{\bBigg@{4}}
\makeatother
\DeclareMathAlphabet\mathbfcal{OMS}{cmsy}{b}{n}
\allowdisplaybreaks

\begin{document}
\title{Third post-Newtonian gravitational radiation from two-body scattering II:\\
Hereditary Energy radiation
}
\date{\today}
\author{Gihyuk Cho}
\email{gihyuk.cho@desy.de}
\affiliation{Deutsches Elektronen-Synchrotron DESY, Notkestr. 85, 22607 Hamburg, Germany}
\begin{abstract}
We compute the hereditary part of the third post-Newtonian accurate gravitational energy radiation
from hyperbolic scatterings (and parabolic scatterings) of non-spinning compact objects. We employ large angular momentum ($j$) expansion, and compute it to the relative $1/j^{11}$ order (so the first 12 terms). For parabolic scattering case, the exact solution is computed. At the end, the completely collected expression of the energy radiation upto the third post-Newtonian and from $1/j^{3}$ to $1/j^{15}$ order, is presented including the instantaneous contribution.
\end{abstract}

\maketitle

\section{Introduction}
Accurate prediction of energy radiation from compact object binaries emitted via gravitational waves (GW), is one of the crucial inputs to predict dynamics of the binary system. The behavior of gravitational field at future null infinity is not just dependent on \textit{instantaneous} dynamics of the binaries  at the same retarded time, but also dependent on the past history of the system due to non-linearity of Einstein's field equation \cite{blanchet2014}. The latter contribution is called \textit{hereditary} \cite{Blanchet:1987wq,Blanchet:1992br,Blanchet:1993ec} understood within the framework of post-Newtonian (PN) theory and multipolar post-Minkwoskian (MPM) expansion \cite{Blanchet:1985sp, Blanchet:1989ki} and effective field theory approach \cite{Galley:2015kus}.
Since more attention has been paid on bound orbits \cite{abbott2017gw170817,abbott2017multi, abbott2019gwtc,abbott2021gwtc,venumadhav2020new} in GW/multi-messengers astronomy, energy radiation as well as angular momentum radiation from bound orbits were already computed upto the third post-Newtonian (3PN) order, which indicates $1/c^6$ ($c$ is speed of light) correction to leading order term, in \cite{arun2008inspiralling,arun2009third,Arun:2007rg} long ago in both instantaneous/hereditary contributions. On the other hand, highly accurate description on unbound orbits is getting more attention today in several complementary methodologies \cite{Cho:2018,thefirst,Kalin:2020lmz,Dlapa:2021vgp,Dlapa:2021npj,Liu:2021zxr}, because of its direct usages such as for detection in LIGO, LISA and IPTA \cite{garcia2018gravitational,mukherjee2020detectability,kocsis2006detection,burke2019astrophysics}, and its astronomical applications to gravitational captures \cite{bae2020},  and also the fact that analytic solutions describing dynamics of unbound orbits have comprehensive information on both bound and unbound orbits \cite{Kalin:2019rwq,Kalin:2019inp,Cho:2021arx}.  \\ 
In this work, we compute the hereditary contribution of energy radiation ($\Delta {\cal E} $) during hyperbolic and parabolic scattering of non-spinning compact objects upto 3PN order. This is a sequel of Ref.~\cite{thefirst}, where 3PN instantaneous contribution of $\Delta {\cal E} $ and angular momentum  radiation ($\Delta {\cal J} $) were computed. By Ref.~\cite{thefirst} and this work, we complete 3PN energy radiation. Historically, the leading (Newtonian) order energy radiation was computed in \cite{Hansen:1972jt}, and the extension to 1PN order was made in \cite{blanchet1989}. And post-Minkowskian (PM) result was computed at the leading $G^3$ order in \cite{Herrmann:2021} ($G$ is Newton's constant).
Recently, the hereditary contribution of both $\Delta {\cal E} $ and $\Delta {\cal J} $  was tackled in \cite{Bini:2021qvf} partially upto  $1/j^7$ order, at the leading order of each tail, tail-of-tail, tail-squared pieces. In independent way, we compute all pieces of the hereditary contribution upto 3PN order including next-leading order of tail piece, and provide upto $1/j^{15}$ or $1/j^{16}$ order in large-$j$ (or, large eccentricity) expansion. Since taking large angular momentum approximation after PN approximation sequentially is equivalent to small energy approximation after PM expansion, this is corresponding to compute upto $G^{15}$ (for tail), and $G^{16}$ (for tail-of-tail, tail-squared) orders. The reason that the aimed $G$ order is different is that our purpose is not only providing a truncated solution in a certain $G$ order, but provide a decent approximation with 12 terms for practical usages. Since tail piece starts at $G^{4}$ order, while tail-of-tail and tail-squared pieces start at $G^{5}$ order, they could end at the different orders. Also, for this reason, we work with two variables, angular momentum and eccentricity to characterize hyperbolic orbits. We will see that this choice exhibits remarkable convergence in Sec.\ref{sec:parab}. 
\par
This paper is organized as what follows. In Sec.\ref{sec:formalism}, we explain all inputs needed for entire computation. Next, we do the main computation. The three different pieces are solved via several strategies adopted for each pieces, leading order of tail in Sec.\ref{sec:tail}, leading order of tail-of-tail, tail-squared in Sec.\ref{sec:tails} and 1PN correction of tail in Sec.\ref{sec:1PNtail}. In Sec.\ref{sec:parab}, we provide exact value of energy radiation from parabolic orbits, and it will be compared to hyperbolic energy radiation. In Appendix.\ref{elementaryintegral}, some required integrals are computed exactly. In Appendix.\ref{sec:fullresult}, we collect all contributions upto 3PN and from $G^3$ to $G^{15}$ including the instantaneous part. In the ancillary file accompanied with this paper, every results are provided in \textit{Wolfram} language.
 
\section{Formalism \& Notations}\label{sec:formalism}
We are computing total energy radiation from scatterings of compact binaries of which each components of which masses are $m_1$, $m_2$, and their positions are $x_1^i$, $x_2^i$ in an almost Minkowskian coordinate. Since total energy radiation $\Delta\mathcal{E}$ is composed of several pieces schematically,
\begin{align}
\Delta\mathcal{E}=\Delta\mathcal{E}_\text{inst}+\Delta\mathcal{E}_\text{hered}\,,
\end{align}
where $\Delta\mathcal{E}_\text{hered}$, of our interest, is consist with sub-pieces,
\begin{align}
\Delta\mathcal{E}_\text{hered}=\Delta\mathcal{E}_\text{tail}+\Delta\mathcal{E}_{(\text{tail})^2}+\Delta\mathcal{E}_\text{tail(tail)}\,.
\end{align}
Since $\Delta\mathcal{E}_\text{tail}$ starts at 1.5PN order, we need the leading term and its 1PN correction (2.5PN), while only leading contribution of $\Delta\mathcal{E}_{(\text{tail})^2}+\Delta\mathcal{E}_\text{tail(tail)}$ is required, because they start at 3PN order. To do this, we need 1PN quadrupole mass moment $I_{ij}$, Newtonian octupolar mass moment $I_{ijk}$ and Newtonian quadrupole current moment $J_{ij}$ in terms of relative position $x^i=x_1^i-x_2^i$ of binaries in the center of mass frame, which can be found in \cite{arun2008inspiralling}, and we present them here for convenience, 
\begin{widetext}
\begin{subequations}\label{moments}
\begin{align}
I_{ij}&=M\,\nu\,\Big(x_i\,x_j\Big)_\text{STF}\,\Bigg[1+v^2\,\Big(\frac{29}{42}-\frac{29}{14}\nu\Big)+\frac{G\,M}{R}\Big(-\frac{5}{7}+\frac{8}{7}\nu\Big)\Bigg]\notag\\
&+\frac{M\,\nu}{c^2}\,\Big(v_i\,v_j\Big)_\text{STF} \,R^2\,\Big(\frac{11}{21}-\frac{11}{7}\,\nu\Big)+\frac{M\,\nu}{c^2}\,\Big(x_i\,v_j\Big)_\text{STF}\,\frac{d(R^2)}{d\hat{t}}\,\Big(-\frac{2}{7}+\frac{6}{7}\,\nu\Big) \,,\\[1ex]
I_{ijk}&=-M\,\nu\,\sqrt{1-4\,\nu}\,\Big(x_i\,x_j\,x_k\Big)_\text{STF}\,,\\[1ex]
J_{ij}&=M\,\nu\,\sqrt{1-4\,\nu}\,\,\Big(x^i (x\times v)^j\Big)_\text{STF}\,.
\end{align}
\end{subequations}
Therein, the followings are defined : $v^i=\frac{dx^i}{d\hat{t}}$ where $\hat{t}$ stands for the physical time, total mass $M=m_1+m_2$, symmetric mass ratio $\nu=\frac{m_1\,m_2}{M^2}$,  $R=\sqrt{x^ix^j\delta_{ij}}$, and  $()_\text{STF}$ means the symmetric and trace-free (STF) part.  And $x^i$ in terms of physical time $\hat{t}$ upto 3PN can be found in Ref.~\cite{Cho:2018} (we need only 1PN), and is written here with introducing reduced variables such as $r:=\frac{R}{G\,M}$ and $t:=\frac{t_\text{phys}}{G\,M}$. Note that $r$ has a dimension of $\frac{\text{time}^2}{\text{length}^2}$ while $t$ has $\frac{\text{time}^3}{\text{length}^3}$. And we define the azimuthal angle as $\phi:=\arctan(\frac{x^2}{x^1})$ assuming that $x^3=0$. The analytic expressions for these dynamical variables are given in time implicitly via eccentric anomaly $u$ ($-\infty\leq u\leq\infty$) characterized by $e_t$ (time eccentricity) in 1PN relation, 
\begin{subequations}\label{qk}
\begin{align}
r&=j^2\,\frac{e_t\,\cosh u-1}{e_t^2-1}+\frac{1}{c^2}\,\frac{8+4\,e_t^2\,(-3+\nu)-e_t\,\big(e_t^2\,(-4+\nu)+3\,\nu\big)\,\cosh u}{2\,(e_t^2-1)}\,,\\[1ex]
\phi&=2 \arctan\left(\sqrt{\frac{e_t+1}{e_t-1}} \tanh \left(\frac{u}{2}\right)\right)+\frac{1}{c^2\,j^2} \left(6 \arctan\left(\sqrt{\frac{e_t+1}{e_t-1}} \tanh \left(\frac{u}{2}\right)\right)-\frac{e_t \sqrt{e_t^2-1} \,(-4+\nu ) \sinh (u)}{
   (e_t\, \cosh (u)-1)}\right)\,,\\[1ex]
   t&=\frac{1}{n}(e_t\,\sinh u-u)\,,
\end{align}
\end{subequations}
where
\begin{align}
n=\frac{(e_t^2-1)^{3/2}}{j^3}+\frac{1}{c^2\,j^5}\,\frac{(e_t^2-1)^{3/2}\,(3+\nu+e_t^2\,(-9+5\,\nu))}{2}\,,
\end{align}
and $e_t$ is dimensionless and given in reduced energy $E=\frac{\mathcal{E}}{M\,\nu}$, reduced angular momentum $j=\frac{\mathcal{J}}{G\,M^2\,\nu}$ as,
\begin{align}
e_t^2=1+2\,E\,j^2+\frac{E}{2\,c^2}\,\Big(8-8\,\nu+(17-7\,\nu)(2\,E\,j^2)\Big)\,.
\end{align} 
Note that we will adopt $e_t$ and $j$ to characterize hyperbolic orbits, because $e_t$ gives the simplest structure in $t$ which makes the following computations easier, and also large-$e_t$ expansion (which is equivalent to large-$j$ expansion) with $j$ exhibits great convergence rate even at $e_t=1$ as reported in \cite{thefirst}.
\end{widetext}
\section{Leading order Tail Contribution}\label{sec:tail}
In this section, we are going to get the analytic expressions of the Newtonian contribution of each tail contributions of energy radiation. The computation done in this section is influenced by \cite{Bini:2017wfr,Bini:2020hmy,Bini:2020rzn}. The tail contribution up to 3PN order, the explicit expressions \cite{Arun:2007rg} are given as 
\begin{widetext}
\begin{align}
\Delta\cE_\text{tail} &=\frac{1}{c^8}\,\frac{4\,G^2\,\cM}{5}\,\frac{1}{G^6\,M^6} \int^\infty_{-\infty} \,dt \,\int^\infty_0\,d\tau\, I^{(3)}_{ij}(t)\,I^{(5)}_{ij}(t-\tau)\,\Bigg[\,\log\Big(\frac{c\,\tau \,G\,M}{2\,r_0}\Big)+\frac{11}{12}\Bigg]\nl
&+\frac{1}{c^{10}}\frac{4\,G^2\,\cM}{189}\,\frac{1}{G^8\,M^8} \int^\infty_{-\infty} \,dt \,\int^\infty_0\,d\tau\, I^{(4)}_{ijk}(t)\,I^{(6)}_{ijk}(t-\tau)\,\Bigg[\,\log\Big(\frac{c\,\tau \,G\,M}{2\,r_0}\Big)+\frac{97}{60}\Bigg]\nl
&+\frac{1}{c^{10}}\frac{64\,G^2\,\cM}{45}\,\frac{1}{G^6\,M^6} \int^\infty_{-\infty} \,dt \,\int^\infty_0\,d\tau\, J^{(3)}_{ij}(t)\,J^{(5)}_{ij}(t-\tau)\,\Bigg[\,\log\Big(\frac{c\,\tau\,G\,M}{2\,r_0}\Big)+\frac{7}{6}\Bigg]\,,
\end{align}
\end{widetext}
where $\cM$ is ADM mass
\begin{align}
\cM=M\,\Bigg(1+\frac{\nu}{c^2}\, \frac{e_t^2-1}{2\,j^2}\Bigg)\,.
\end{align}
Note that the extra $G\,M$s in the above expressions, which cannot be found in \cite{Arun:2007rg} appear because of using $t$ and $\tau$ as the reduced time variable. In favor of computation, we decompose $\Delta\cE_\text{tail}$ into several pieces in Fourier domain such as
\begin{subequations}\label{decomposition}
\begin{align}
&\Delta\cE_{(0)}+\Delta\cE_{(1)} =\frac{2}{5}\frac{n^6\,\cM}{c^8\,G^4\,M^6\,e_t^6}\int^\infty_0 dp\,|\hat{I}_{ij}(p)|^2\, p^7\,,\\[1ex]
&\Delta\cE_{(2)} =\frac{2}{189}\frac{n^8\,\cM}{c^{10}\,G^6\,M^8\,e_t^8}\, \int^\infty_0 \,dp\,|\hat{I}_{ijk}(p)|^2\,p^9\,,\\[1ex]
&\Delta\cE_{(3)}=\frac{32}{45}\,\frac{n^6\,\cM}{c^{10}\,G^4\,M^6\,e_t^6}\, \int^\infty_0 \,dp\,|\hat{J}_{ij}(p)|^2\,p^7\,,
\end{align}
\end{subequations}
where $\Delta\cE_{(0)}$, $\Delta\cE_{(2)}$ and $\Delta\cE_{(3)}$ denote the leading(Newtonian) order contributions of the right hand sides respectively, whereas $\Delta\cE_{(1)}$ refers to 1PN correction of the right hand side.
Therein, we have defined the Fourier transform as
\begin{subequations}
\begin{align}
\hat{K}(p)&:=\int^{+\infty}_{-\infty}d\bar{l}\, \text{e}^{i\,p\,\bar{l}} K(\bar{l})\,,\\[1ex]
K(t)&=\frac{1}{2\,\pi}\int^{+\infty}_{-\infty}dp\, \text{e}^{-i\,p\,\bar{l}}\hat{K}(p)\,,
\end{align}
\end{subequations}
where the reduced mean anomaly is defined as $\bar{l}:=\sinh u-\frac{u}{e_t}=\frac{n}{e_t}\,(t-t_0)$ in the preparation of large $e_t$ expansion, and $p$ and $l$ are both dimensionless for the convenience of forthcoming computation. Also, we should mention that we use the following integration reported in \cite{Arun:2007rg} (with $\gamma$ being Euler's constant), 
\begin{align}\label{log1}
\int^{+\infty}_0 dt\,\ex^{i\,\sigma\,t}\log (t)=-\frac{1}{\sigma}\,\Bigg[\frac{\pi}{2}\,\text{sign}(\sigma)+i\,\Big(\log\big(|\sigma|\big)+\gamma\Big)\Bigg]
\end{align} 
in the middle of obtaining the Fourier domain expressions in Eqs.~(\ref{decomposition}). \\
Among them, we compute three of them $\Delta\cE_{(0)}$, $\Delta\cE_{(2)}$ and $\Delta\cE_{(3)}$ here, because they have analytic expressions in Fourier domain, whereas it is not the case for $\Delta\cE_{(1)}$. We will come back to $\Delta\cE_{(1)}$ in Sec.~\ref{sec:1PNtail}.\\

Using Eqs.~(\ref{qk}), we can write all components of multipole moments in terms of the eccentric anomaly $u$ (particularly  the form of $\ex^{k u}$ with integers $k$).
By considering $d\bar{l}=du (\cosh u-\frac{1}{e_t})$, every terms encountered in the Fourier transformation could be written in form of the typical integral,
\begin{align}\label{typicalintegral}
I_t:=\int^{+\infty}_{-\infty} du \,\ex^{i\,p\,\sinh u+k \,u-i\,\frac{p}{e_t}\,u}\,.
\end{align}
This is nothing but one of integral representations of Hankel function of the first kind $H^{(1)}_{a}(x)$ (we will call it just the Hankel function henceforth),
\begin{align}
H^{(1)}_{a} (x)=\frac{1}{i\,\,\pi}\int ^{+\infty}_{-\infty}\,\text{e}^{x \sinh u-a u} \,du\,.
\end{align} 
Note that the integral on the right hand side is only convergent when the order is in $-1<\text{Re}(a)<1$, which is not our case  ($k$ could be bigger than 1 in $I_t$). It implies that the typical integral $I_t$ is actually divergent in the usual integral sense. But the Hankel function is even analytic outside of $-1<\text{Re}(a)<1$, hence, we will use the Hankel function as a regularized, hence finite value of the typical divergent integral by the argument of analytic continuation resulting in
\begin{align}
I_t=i\,\pi\,H^{(1)}_{i\,\frac{p}{e_t}-k}(i\,p).
\end{align}
After successive uses of the relations 
\begin{align}
H^{(1)}_{a-1}(x)+H^{(1)}_{a+1}(x)=\frac{2a}{x}H^{(1)}_{a}(x)\,,
\end{align}
to unify all the orders appearing in the integrands to either $0$ or $1$, we obtain the intermediate results of $|\hat{I}_{ij}(p)|^2$, $|\hat{I}_{ijk}(p)|^2$ and $|\hat{J}_{ij}(p)|^2$ at the leading order.
As pointed out in \cite{Bini:2017wfr}, the large-$e_t$ expansion drops the $p$ dependence in the order of the Hankel function, which makes the computation much simpler. After taking the large-$e_t$ expansion, every terms can be dealt by the following master integral (similar expression is reported in Eq.(7.24) of \cite{Bini:2020rzn}),
\begin{widetext}
\begin{align}\label{masterfourier}
\int ^{+\infty}_{-\infty} \,dp\,p^n H^{(1)}_{a}(i\,p)\,H^{(1)}_{b} (i\,p)=& \frac{2^n\,\text{e}^{-i\frac{(1+a+b)\,\pi}{2}}}{i\,\,\pi\,\Gamma(1+n)} \,\Gamma\Big(\frac{1-a-b+n}{2}\Big)\,\Gamma\Big(\frac{1+a-b+n}{2}\Big)\nl
&\times\Gamma\Big(\frac{1-a+b+n}{2}\Big)\,\Gamma\Big(\frac{1+a+b+n}{2}\Big)\,,
\end{align}
\end{widetext}
Now that all the computation is reduced down to the straightforward but extensive machinery work, we only present the final results with the first 12 terms in the large $e_t$ expansion $\Delta\cE^{(l)}_{(0)}$ in Appendix.\ref{sec:partialresult}. (We use the superscript $(l)$ to denote that this is not exact one but expressed in large-$e_t$ (or, large-$j$) expansion.) We have found that the first three terms are coincident to Eq.(4.14) in \cite{Bini:2021qvf}. By the exactly same procedure, we also get the first 12 terms of $\Delta\cE_{(2)}$ and $\Delta\cE_{(3)}$ presented in Appendix.\ref{sec:partialresult}.
\section{Leading Quadratic tail contribution}\label{sec:tails}
In this section, we are going to obtain analytic expressions of $(\text{tail})^2$ and $\text{tail}(\text{tail})$ contributions. In terms of mulitpole moments, the leading order terms of each pieces \cite{Arun:2007rg} are
\begin{widetext}
\begin{subequations}
\begin{align}
\Delta\cE_{(\text{tail})^2}&=\frac{4\,G^3\,\cM^2}{c^{11}\,5}\,\frac{1}{G^7\,M^7}\int^{+\infty}_{-\infty}\,dt\,\Bigg(\int^{+\infty}_0\,d\tau\,I^{(5)}_{ij}(t-\tau)\Bigg[\log\Big(\frac{c\,\tau\,G\,M}{2\,r_0}\Big)+\frac{11}{12}\Bigg]\Bigg)^2\,,\\[1ex]
\Delta\cE_{\text{tail}(\text{tail})}&=\frac{4\,G^3\,\cM^2}{c^{11}\,5}\,\frac{1}{G^7\,M^7}\int^{+\infty}_{-\infty}\, dt\, I^{(3)}_{ij}(t)\int^{+\infty}_{0} \,d\tau\,I^{(6)}_{ij}(t-\tau)\,\Bigg[\log^2\Big(\frac{c\,\tau\,G\,M}{2\,r_0}\Big)+\frac{57}{70}\,\log\Big(\frac{c\,\tau\,G\,M}{2\,r_0}\Big)+\frac{124627}{44100}\Bigg]\,.
\end{align}
\end{subequations}
Similarly, they are rearranged compactly in Fourier domain into
\begin{subequations}
\begin{align}
\Delta\cE_{(4)}&=\frac{\cM^2}{c^{11}\,G^4\,M^7}\,\frac{2 }{5\,\pi}\,\frac{n^7}{e_t^7}\,\Bigg[\frac{2\,\pi^2}{3}-\frac{214}{105}\Bigg(\,\log \Big(\frac{2\, n \,r_0}{c\,e_t\,G\,M}\Big)+\gamma_E\Bigg)-\frac{116761}{29400}\Bigg]\,\Big( \int^\infty_0 dp\,|\hat{I}_{ij}|^2\,p^8\Big)\,,\\[2ex]
\Delta\cE_{(5)}&=-\frac{\cM^2}{c^{11}\,G^4\,M^7}\,\frac{428}{525\,\pi}\,\frac{n^7}{e_t^7}\, \int^\infty_0 dp\,|\hat{I}_{ij}|^2\,p^8\,\log p\,,
\end{align}
\end{subequations}
where we use the relations \cite{Arun:2007rg}, Eq.(\ref{log1}) and
\begin{align}\label{log2}
\int^{+\infty}_0 dt\,\ex^{i\,\sigma\,t}\log^2 (t)=\frac{i}{\sigma}\,\Bigg\{\frac{\pi^2}{6}-\Bigg[\frac{\pi}{2}\,\text{sign}(\sigma)+i\,\Big(\log\big(|\sigma|\big)+\gamma\Big)\Bigg]^2\Bigg\}\,.
\end{align} 
Thanks to even power of $p$ in the integrand, $\Delta\cE_{(4)}$ can be integrated in an exact way. Using the elementary integrals $E^i(n)$ defined in Appendix.\ref{elementaryintegral},
\begin{align}
&\int^\infty_0 dp\,|\hat{I}_{ij}|^2\,p^8=8\,\pi^2\,e_t^3\, j^8\,G^4\,M^6\,\nu^2\nl
&\times\Bigg[\frac{\left(3 \,e_t^2-4\right)\, e_t\, E^{y}(5)}{\left(e_t^2-1\right)^3}+\frac{e_t^2\,E^{z}(4)}{\left(e_t^2-1\right)^3}+\frac{e_t^2\,E^{z}(6)}{\left(e_t^2-1\right)^2}-\frac{E^{x}(6)}{e_t^2-1}-\frac{\left(e_t^4-3\, e_t^2+3\right)\, E^{x}(4)}{3 \,\left(e_t^2-1\right)^4}\Bigg]\,,
\end{align}
where the exact solutions of each elementary integrals are listed in Eqs.(\ref{eleIntegrals}).\end{widetext}
Gathering all pieces, we obtain the exact expression ($\cM=M$ at this order) displayed in Eq.(\ref{deltaE4}) of Appendix.\ref{sec:partialresult}. We have checked that the $r_0$ dependence in $\Delta \cE_{(4)}$ is canceled with $r_0$ in Eq.(31.d)~ of \cite{thefirst}, so the final result does not depend on $r_0$.\\
In the case of $\Delta\cE_{(5)}$, since there is no closed form solution, we use the same strategy of computing the Newtonian tails explained in Sec.~\ref{sec:tail}. The picky $\,\log p$ can be dealt by the derivative with respect to $n$ in the master integral Eq.(\ref{masterfourier}), and hence is reduced down to the machinery work as well. Likewise, we display the first 12 terms of the large-$e_t$ expansion of $\Delta\cE_{(5)}$ in Appendix.\ref{sec:partialresult}. And the agreement is found in the first three terms with Eq.(4.14) in Ref.\cite{Bini:2021qvf}.\footnote{Only the second arxiv version of \cite{Bini:2021qvf} for now.}
\section{1PN tail contribution}\label{sec:1PNtail}
Now, we need to tackle down the most tricky computation of $\Delta\cE_{(1)}$ in this paper. Since we could not obtain Fourier transform of the quadrupole moment at the 1PN order in general, time domain integration is attempted. Contrary to the way we have computed, we could not find the systematic and simple automatic way of integration \textit{i.e.} master integral method, so the integrations should be treated in a rather heuristic way.  For the first 4 terms, we could find exact values of the coefficients by symbolic integration, but we failed for the other 8 terms. Instead, numerical integration and PSLQ algorithm \cite{PSLQ} are adopted to find analytic form of the coefficients. Since we have encountered more than 10 thousands terms, we will explain the process of computation roughly.  \\
By putting all dynamical variables Eq.(\ref{qk}) in the quadrupole moment Eq.(\ref{moments}a), we construct the integral over $u$ and $v$ which are given via the relations to the time variables
\begin{subequations}
\begin{align}
t&= \frac{1}{n}\big(e_t\sinh u-u\big)\,,\\
t-\tau&=\frac{1}{n}\big(e_t\sinh v-v\big)\,.
\end{align}
\end{subequations}
For the efficient computation, we subsequently choose two dynamical parameters $(\kappa,\,q)$ in the integrand,
\begin{subequations}
\begin{align}
\kappa&=\ex^{u}\,,\\
    q&= \frac{\ex^{v}}{\kappa}\,.
\end{align}
\end{subequations}
Schematically, the tail radiation in time domain has the following form
\begin{subequations}
\begin{align}
\Delta\cE
&\sim\int^{+\infty}_{-\infty}dt \int^{t}_{-\infty}d(t-\tau) \,\mathcal{I}_1(t,t-\tau)\,,\\[1ex]
&\sim\int^{+\infty}_{-\infty}du \int^{u}_{-\infty}dv \,\mathcal{I}_2(u,v)\,,\\[1ex]
&\sim\int^{+\infty}_{0}d\kappa \int^{1}_{0}dq \,\mathcal{I}_3(\kappa,q)\,,
\end{align}
\end{subequations}
where $\mathcal{I}_{1,2,3}$ represent some integrands. One can easily see that the choice of $q$ makes two integrals independent. All terms we encounter is classified into four categories,
\begin{subequations}
\begin{align}
I_1&=\frac{(q\,\kappa^2+1)^n\,\kappa^l}{(q-1)^a\,(q^2\,\kappa^2+1)^b}\,,\\[1ex]
I_2&=\frac{(q\,\kappa^2+1)^n\,\,\kappa^l \,\arctan(\frac{\kappa-1}{\kappa+1})}{(q-1)^a\,(q^2\,\kappa^2+1)^b}\,,\\[1ex]
I_3&=\frac{(q\,\kappa^2+1)^n\,\,\kappa^l\,\log^m(q)}{(q-1)^a\,(q^2\,\kappa^2+1)^c}\,,\\[1ex]
I_4&=\frac{(q\,\kappa^2+1)^n\,\,\kappa^l\,\arctan(\frac{\kappa-1}{\kappa+1})\,\log^m(q)}{(q-1)^a\,(q^2\,\kappa^2+1)^b}\,,
\end{align}
\end{subequations}
where $m,l,a,b,c$ denote some positive integers whereas $n$ is an integer.
To solve them, we have found that
\begin{widetext}
\begin{align}
&\int^1_0\, dq\, \frac{\log ^n(q)\,q^m}{(q+a)(q+b)(q+c)(q+d)}\\[1.5ex]
= &(-1)^{n+m+1}\,n!\,\Bigg[\frac{\,a^m\,\text{Li}_{n+1}(\frac{-1}{a})}{(a-b)(a-c)(a-d)}+\frac{b^m\,\text{Li}_{n+1}(\frac{-1}{b})}{(b-a)(b-c)(b-d)}+\frac{c^m\,\text{Li}_{n+1}(\frac{-1}{c})}{(c-a)(c-b)(c-d)}+\frac{d^m\,\text{Li}_{n+1}(\frac{-1}{d})}{(d-a)(d-b)(d-c)}\Bigg]\,,\notag
\end{align}
\end{widetext}
where $n,m$ are positive integer with $m\leq3$, and $\text{Li}_n$ is poly-logarithm. After taking successive derivatives with respect to $a,b,c,d$, we manage to integrate all types of integrand over $q$.  In the cases of $I_1$, $I_2$ and $I_3$, further integration over $\kappa$ is also possible analytically (hence, every coefficients of $\nu^3$ term are determined exactly). However, in the case of $I_4$, we encounter multi-logarithm terms such as
\begin{align}
\int \,dq\, I_4\sim g(Q)\,\arctan(Q)\,\text{Li}_{n}(f(Q))\,,
\end{align}
where $f(Q), g(Q)$ are some rational function of polynomial of $Q$ with $Q=\frac{\kappa-1}{\kappa+1}$.
Given that $\arctan$ is also a logarithmic function, we need to do integration over two or more logarithm. When $n=2$ (\textit{i.e.} dilogarithm),  we could get rid of $\text{Li}_2$ by integration-by-part, but when $n\geq 3$, because the derivative of $\text{Li}_n$ is still a poly-logarithm,  we could not find an analytic way of integration. Instead, for remaining terms, we integrate $I_4$ type integrals ($q$ is already integrated), numerically with $600\sim1000$ digits, and attempt to find their analytic forms using PSLQ algorithm. Since PSLQ algorithm is used for guessing a relation between integers, we need to make a list of candidates. For even orders $e_t^n$ ($n=8,6,4\cdots$), the candidate list is consist with combinations of even powers of $\pi$, and  $\zeta(m)$ with odd integer $m$, such as
\begin{align}
\{1,\pi^2,\pi^4,\cdots,\zeta(3),\pi^2\zeta(3),\cdots\zeta(5),\pi^2\zeta(5),\cdots\}\,,
\end{align}
and odd powers of $\pi$ for odd powers of $e_t$,
\begin{align}
\{\pi,\pi^3,\pi^5\cdots\}\,.
\end{align}
The higher order of $e_t$ we take, the longer candidate lists are required.
We first guess analytic forms of numerical coefficients using first $400\sim800$ digits, and check if the number of digits used in guessed integers is much smaller than $400\sim800$. For example, if from this 40 digits number, $$46.50941502044973026321447260065209280334\,,$$ PSLQ guesses these two long integers (in each numerator and denominator) in totally 55 digits, $$\frac{4320335101191349704192837044}{291827644122831121457024899}\pi\,,$$ we regard it as a failure, because it just stores 40 digits with 55 digits.
But if PSLQ find two digits such as $\frac{3}{2}\pi^3$, we regard it trustworthy. After meeting this standard, we also check if they can predict the next 200 digits, and if so, we confirm them. Using the coefficients determined in this way, we compute the first 12 terms of $\Delta\cE_{(1)}$ which can be found in Appendix.\ref{sec:partialresult}.
\section{Radiation from Parabolic orbits}\label{sec:parab}
In this section, we are going to compute hereditary part of energy radiation $\Delta\cE^{(p)}_\text{hered}$ from parabolic orbits and this could be understood as the limit of $e_t\to1$ (or, $E\to0$) in the hyperbolic radiation. As pointed out in \cite{thefirst}, the choice of the parametrization $(e_t, h)$ (or $(e, h)$) of orbits allows us quickly converging large eccentricity expansions even approximating well at $e_t=1$. We check if this works even in the hereditary contribution as well as the validity of our hyperbolic computations.\\
The parabolic version of keplerian parametrization can be derived by changing $u$ to $w$ via $u=(e_t^2-1)^{1/2} \,w$ in Eqs.(\ref{qk}). After taking the limit of $e_t\to1$, one will get
\begin{widetext}
\begin{subequations}
\begin{align}
r&=\frac{j^2 w^2}{2}+\frac{j^2}{2}+\frac{1}{c^2}\, \left(\frac{\nu -6}{2}+(1-\nu ) w^2\right)\,,\\[1ex]
t&=\frac{j^3 w^3}{6}+\frac{j^3 w}{2}+\frac{j}{c^2}\, \left(-\frac{1}{2}  (\nu -1) w^3-\frac{3}{2} (\nu -1) w\right)\,,\\[1ex]
\phi&=2 \arctan(w)+\frac{2 \left(3\, w^2\, \arctan(w)-\nu\,  w+4 \,w+3 \,\arctan(w)\right)}{c^2\,j^2\, \left(1+w^2\right)}\,.
\end{align}
\end{subequations}
\end{widetext}
By these, it is possible to construct every mulipole moments in terms of $w$. In the computation of $\Delta\cE^{(p)}_{(0)}$ and $\Delta\cE^{(p)}_{(1)}$, the most of the integrations can be done straightforwardly except
\begin{align}
I_p:=\int^{\infty}_{-\infty}dw\,\frac{w \left(w^6+6 w^4-18 w^2+4\right) \arctan(w)}{ \left(w^2+1\right)^7 \left(w^2+4\right)^{7/2}}\,,
\end{align}
which arises in 1PN order of quadrupole tail part. After inserting the integral representation of $$\arctan( w)=\int^{1}_0{dk}\frac{w}{1+w^2\,k^2}\,,$$ the successive integrations over $w$ and $k$ could yield the analytic result. By simplifying the result with the help of the second order of Clausen function, we managed to obtain
\begin{align}
I_p =\frac{385\, \text{Cl}_2\left(\frac{\pi }{3}\right)}{1024\, \sqrt{3}}-\frac{5221}{115200}-\frac{501151 \,\pi }{2799360\, \sqrt{3}}+\frac{77 \,\pi\,  \log (3)}{1024\, \sqrt{3}}\,,
\end{align}
where
\begin{align}
\text{Cl}_2\left(z\right):=-\int^{z}_0\,dx\,\log\Big|2\sin\frac{x}{2}\Big|.
\end{align}
Specially, $\text{Cl}_2\left(\frac{\pi }{3}\right)$ is the maximum value of $\text{Cl}_2\left(z\right)$ when $z$ is real.
\begin{widetext}
As a result, we obtain the quadrupolar tail piece of energy radiation from parabolic orbits $\Delta\cE^{(p)}$\,,
\begin{align}
\Delta\cE^{(p)}_{(0,1)}=\frac{M\,\nu^2\,\pi}{c^{8}\,j^{10}}\vast[\frac{169984}{45 \,\sqrt{3} }+\frac{1}{c^2\,j^{2}}  \left(27720  \, \text{Cl}_2\left(\frac{\pi }{3}\right)+\frac{7949344\,\pi}{1215}-\frac{41631152  }{525 \,\sqrt{3}}+5544\, \pi\, \log (3)+\frac{2818048 \, \nu }{63 \sqrt{3}}\right)\vast]\,.
\end{align}
\end{widetext}
In the case of contributions $\Delta\cE^{(p)}_{(2)}$, $\Delta\cE^{(p)}_{(3)}$, all computation is trivial. And for $\Delta\cE^{(p)}_{(4)}$, it is easily obtained by putting 1 in $e_t$ in Eq.(\ref{deltaE4}) because it is exact,
\begin{subequations}
\begin{align}
\Delta\cE^{(p)}_{(2)}&=\frac{M\,\nu^2\,(1-4\,\nu)\,\pi}{c^{10}\,j^{12}}\frac{1448960}{63 \,\sqrt{3}}\,,\\
\Delta\cE^{(p)}_{(3)}&=\frac{M\,\nu^2\,(1-4\,\nu)\,\pi}{c^{10}\,j^{12}}\,\frac{4096}{9\, \sqrt{3}}\,,
\end{align}
\begin{align}
\Delta\cE^{(p)}_{(4)}&=\frac{M\,\nu^2\,\pi}{c^{11}\,j^{13}}\,\Bigg[-\frac{161249 \,  \gamma}{15}  -\frac{161249}{15} \,  \log \left(\frac{2\,r_0\,n}{e_t\,G\,M\,c}\right)\notag\\[1ex]
&+\frac{10549 \,\pi^2}{3}-\frac{175958827  }{8400}\Bigg]\,,
\end{align}
\end{subequations}
where we keep $\log(n)$, because it is divergent at $e_t=1$, which will be canceled by $\Delta\cE_{(5)}$. \\
The fifth piece $\Delta\cE^{(p)}_{(5)}$ is left. After performing inverse Fourier transform, $\Delta\cE_{(5)}$ is re-expressed in time domain as
\begin{align}
&\Delta\cE_{(5)}=\frac{428}{525}\Bigg[\Big(\gamma+\log\left(\frac{n}{c^3\,e_t}\right)\Big)\,\int^{+\infty}_{-\infty} dt\,\Big(I_{ij}^{(4)}(t)\Big)^2\,\notag\\[1ex]
&+\int^{+\infty}_{-\infty} dt\,\int^{+\infty}_{t} d\tau\, I_{ij}^{(5)}(t+\tau) I_{ij}^{(4)}(\tau)\log(\tau\,c^3)\Bigg]\,.
\end{align}
As previously computed, almost all term can be dealt by elementary integration methods, but $\arctan$ term should be replaced by its integral representation again. One could get 
\begin{align}
&\Delta\cE^{(p)}_{(5)}=\frac{M\,\nu^2\,\pi}{c^{11}\,j^{13}}\Bigg[\frac{161249  \, \gamma }{15}+\frac{161249}{15}\,  \log \left(\frac{n}{c^3\,e_t}\right)\\[1ex]
&-\frac{90210202 }{1575}+\frac{161249}{5} \, \log (2\,c\,j)+\frac{161249}{30}\,  \log (3)\Bigg]\,.\notag
\end{align}
Indeed, the $\log(n)$ term in $\Delta\cE^{(p)}_{(4)}$ and $\Delta\cE^{(p)}_{(5)}$ are canceled. Contrary to $\Delta\cE^{(p)}_{(5)}$ having $\log(j)$ dependence, $\Delta\cE^{(l)}_{(5)}$ does not have it. This missing $\log(j)$ comes in from $\log(n)=\log\left(\frac{(e_t^2-1)^{3/2}}{j^3}\right)$ in $\Delta\cE_{(4)}$ (Eq.(\ref{deltaE4})), which also indicates that $\Delta\cE_{(4)}$ and $\Delta\cE_{(5)}$ should not be treated independently.\\
Finally, we compare the exact parabolic results $\Delta\cE^{(p)}$ and the hyperbolic results in large-$e_t$ expansion $\Delta\cE^{(l)}$ at $e_t=1$ in Table.\ref{table:1}. One could find that they have very close values, which shows the robustness of our computations as well as the efficiency of large-$e_t$ expansion. 
\begin{table}
\begin{tabular}{|c|c|c|c| } 
\hline
&$\Delta\cE^{(p)}$& $\Delta\cE^{(l)}|_{e_t=1}$  \\
\hline \hline
$\Delta\cE_{(0)}$ &107.0544 & 107.0545 \\ 
$\Delta\cE_{(1)}$ &1240.453 & 1240.384 \\ 
$\Delta\cE_{(2)}+\Delta\cE_{(3)}$ &332.3571& 332.3535 \\ 
 $\Delta\cE_{(4)}+\Delta\cE_{(5)}$&-1114.843 &-1114.842 \\ 
\hline
\end{tabular}
\caption{7 digits comparisons. $G=M=c=j=r_0=1$ and $\nu=1/8$ are chosen}
\label{table:1}
\end{table}\\
For the completeness, we present the completed result of $\Delta\mathcal{E}^{(p)}=\Delta\mathcal{E}^{(p)}_\text{inst}+\Delta\mathcal{E}^{(p)}_\text{hered}$ from parabolic orbits upto 3PN combined with the instantaneous contribution (Eq.(43) of \cite{thefirst}), 
\begin{widetext}
\begin{align}
\Delta\cE^{(p)}&=\frac{M\,\nu^2\,\pi}{c^5\,j^7}\,\vast\{\frac{170}{3}+\frac{1}{c^2\,j^2}\,\Bigg(\frac{13447}{20}-\frac{1127 \,\nu }{3}\Bigg)+\frac{1}{c^3\,j^3}\,\frac{169984}{45\sqrt{3}}+\frac{1}{c^4\,j^4}\,\Bigg(\frac{5839651}{1008}-\frac{258051 \,\nu}{40}+\frac{5481 \,\nu ^2}{4}\Bigg)\nl
&+\frac{1}{c^5\,j^5}\,\Bigg(27720 \,\text{Cl}_2\left(\frac{\,\pi }{3}\right)-\frac{29317552}{525 \sqrt{3}}+\frac{7949344 \,\pi }{1215}+5544 \,\pi  \log (3)-\frac{3092480 \,\nu }{63 \sqrt{3}}\Bigg)\nl
&+\frac{1}{c^6\,j^6}\,\Bigg[\frac{161249 \log (2\,\sqrt{3}\,c\,j)}{15}+\frac{10549 \,\pi^2}{3}+\frac{16880186749}{241920}+\left(-\frac{3972009943}{30240}+\frac{208813 \,\pi^2}{160}\right) \,\nu\nl
&\quad+\frac{1157409 \,\nu^2}{32}-\frac{29645 \,\nu^3}{8}\Bigg]\vast\}+\cO\Bigg(\frac{1}{c^{12}}\Bigg).
\end{align}
\end{widetext}
\section{Discussion \& Future work}
We have computed the hereditary part of energy radiation from hyperbolic scatterings as well as parabolic scatterings. For hyperbolic computation, we rely on large-$e_t$ (or, large-$j$) expansion while the parabolic result is exact.  As expected, the hyperbolic result in large-$e_t$ is close to the value of the parabolic one numerically. It is also interesting to point out that the irrational values appear in the parabolic case is $\pi$, $\log(2)$, $\log(3)$, $\sqrt{3}$ and $\text{Cl}_2(\frac{\pi}{3})$ whereas $\pi$, $\log(2)$, $\gamma$ and $\zeta(2n+1)$ in the hyperbolic case.  In order to complete 3PN radiation reaction to scattering angle, impact parameter, we need to compute angular momentum radiation. This is going to be done in a subsequent paper using the same techniques elaborated in this paper except the memory effect which will probably require another treatment. We also expect that the practical knowledge on the integrations gained from this work, might be helpful to extend our knowledge specially in non-local in time conservative dynamics, which has the similar structure with hereditary radiation.

\section*{Acknowledgments}
We thank Yannick Boetzel for the setup of the formalism at the beginning phase, Rafael A. Porto for discussion and Hyung-Mok Lee for technical support.  G.C is supported by the ERC Consolidator Grant “Precision Gravity: From the LHC to
LISA,” provided by the European Research Council (ERC) under the European Union’s H2020 research and innovation
programme (grant No. 817791). 
\newpage
\begin{widetext}
\appendix
\section{Exact integrals}\label{elementaryintegral}
The aim of this section is to obtain the following element integrals, 
\begin{subequations}
\begin{align}
E^x(n)&:=\int^\infty_0dq\,q^n\,(H^{(1)}_{i\,q}(i \,q\,e_t))^2\,,\\
E^y(n)&:=i\,\int^\infty_0dq\,q^n\,H_{i\,q}(i \,q\,e_t)\,\dot{H}^{(1)}_{i\,q}(i \,q\,e_t)\,,\\
E^z(n)&:=\int^\infty_0dq\,q^n\,(\dot{H}^{(1)}_{i\,q}(i \,q\,e_t))^2\,,
\end{align}
\end{subequations}
where $\dot{H}^{(1)}_a(x)=\frac{\partial}{\partial x}H^{(1)}_a(x)$. Inspired by Appendix of Ref.\cite{Peters:1963ux}, we start at the following relation between $(l,u,e_t)$,
\begin{align}\label{imp}
l= e_t\,\sinh u-u\,.
\end{align}
This gives values of $l$ from $e_t$ and $u$ explicitly. What we want to do is to obtain an explicit expression  $u=U(l,e_t)$.
First, we seek Fourier transform $ K(q)$ of $e\,\sinh u$,
\begin{align}
e_t\,\sinh u:=\int^{+\infty}_{-\infty}\, K(q)\,\text{e}^{-i\, q\, l} \,dq\,,
\end{align}
or 
\begin{align}
K(q)&=\frac{e_t}{2\pi}\,\int \,\sinh u\, \text{e}^{i\, q\, l}dl\,,\notag\\
&=-\frac{H^{(1)}_{i\, q} (i\,q\,e_t)}{2\,q}\,.
\end{align}
So,
\begin{align}\label{exp}
U(l,e_t)=-l-\int^{+\infty}_{-\infty}\, \frac{H^{(1)}_{i\, q} (i\,q\,e_t)}{2\,q}\,\text{e}^{-i\, q\, l} \,dq\,.
\end{align}
Now, we have two definitions for $U$, the implicit one Eq.(\ref{imp}), and the explicit one Eq.(\ref{exp}). If one takes an operator $\nabla_{e_t}:=\partial^2_{e_t}+\frac{1}{{e_t}}\partial_{e_t}$ on the definitions, two seemingly different (but actually the same) expressions are obtained,
\begin{align}
\nabla_{e_t}\,U&=\frac{\left(e_t^2-1\right) \sinh (u)}{e_t \,(e_t \cosh (u)-1)^3}\quad(\text{from the implicit side})\,,\\[1ex]
&=\frac{(e_t^2-1)}{2\,e_t^2}\int^{+\infty}_{-\infty}\, q\,H^{(1)}_{i\, q} (i\,q\,e_t)\,\text{e}^{-i\, q\, l} \,dq\quad(\text{from the explicit side})\,,
\end{align}
in which we apply $$
\dot{H}^{(1)}_a(x)= \frac{a\, H_a^{(1)}(x)}{x}-H_{a+1}^{(1)}(x)\,.
$$ If one repeats this several times, since 
\begin{align}
\big(\nabla_{e_t}\big)^n\,U\sim \int^{+\infty}_{-\infty}\, q^{2n-1}\,H^{(1)}_{i\, q} (i\,q\,e_t)\,\text{e}^{-i\, q\, l} \,dq\,,
\end{align}
from the explicit side, we multiply it with its complex conjugate ($*$) to get the desirable form,
\begin{align}
\Big[\big(\nabla_{e_t}\big)^nU\Big]\Big[\big(\nabla_{e_t}\big)^m\,U\Big]^{*}\sim\int\int dq \,dk \,q^{2n-1}\,k^{2m-1}\,H^{(1)}_{iq}(iqe_t)\,H^{(1)*}_{ik}(ike_t)\text{e}^{-i (q-k)l}\,.
\end{align}
On the other hand, one will get elementary expressions from the implicit side.
After integrating over $-\infty<l<\infty$ on the both sides and considering $H^{(1)}_{iq}(iqe_t)$ is pure imaginary number, one can get exact values of $E^x$ type integrals. In order to get $E^{y}$ and $E^z$ types, taking derivatives with respect to $e_t$ is sufficient. As a result, we list the required integrals below :
\begin{subequations}\label{eleIntegrals}
\begin{align}
E^x(2)&=\frac{13 \,e_t^2+2}{12 \,\pi  \left(\,e_t^2-1\right)^3}+\frac{\left(\,e_t^2+4\right) \,e_t^2 \arccos\left(-\frac{1}{\,e_t}\right)}{4 \,\pi  \left(\,e_t^2-1\right)^{7/2}}\,,\\[1ex]
E^x(4)&=-\frac{3691 \,e_t^6+11082 \,e_t^4+2568 \,e_t^2-16}{960 \,\pi  \left(\,e_t^2-1\right)^6}-\frac{\left(27 \,e_t^6+472 \,e_t^4+592 \,e_t^2+64\right) \,e_t^2 \arccos\left(-\frac{1}{\,e_t}\right)}{64 \,\pi  \left(\,e_t^2-1\right)^{13/2}}\,,\\[1ex]
E^x(6)&=-\frac{4954041 \,e_t^{10}+45033746 \,e_t^8+65383216 \,e_t^6+18070896 \,e_t^4+565696 \,e_t^2+1280}{161280 \,\pi  \left(\,e_t^2-1\right)^9}\nl
&-\frac{\left(1125 \,e_t^{10}+45820 \,e_t^8+189040 \,e_t^6+161152 \,e_t^4+27776 \,e_t^2+512\right) \,e_t^2 \arccos\left(-\frac{1}{\,e_t}\right)}{512 \,\pi  \left(\,e_t^2-1\right)^{19/2}}\,,\\[1ex]
E^y(3)&=\frac{5 \,e_t \left(11 \,e_t^2+10\right)}{24 \,\pi  \left(\,e_t^2-1\right)^4}+\frac{\,e_t \left(3 \,e_t^4+24 \,e_t^2+8\right) \arccos\left(-\frac{1}{\,e_t}\right)}{8 \,\pi  \left(\,e_t^2-1\right)^{9/2}}\,,\\[1ex]
E^y(5)&=\frac{\,e_t \left(7517 \,e_t^6+39294 \,e_t^4+26296 \,e_t^2+1968\right)}{640 \,\pi  \left(\,e_t^2-1\right)^7}+\frac{\,e_t \left(135 \,e_t^8+3520 \,e_t^6+8160 \,e_t^4+3072 \,e_t^2+128\right) \arccos\left(-\frac{1}{\,e_t}\right)}{128 \,\pi  \left(\,e_t^2-1\right)^{15/2}}\,,\\[1ex]
E^z(4)&=\frac{5469 \,e_t^6+12598 \,e_t^4+2392 \,e_t^2+16}{960 \,\pi  \,e_t^2 \left(\,e_t^2-1\right)^5}+\frac{\left(45 \,e_t^6+632 \,e_t^4+624 \,e_t^2+64\right) \arccos\left(-\frac{1}{\,e_t}\right)}{64 \,\pi  \left(\,e_t^2-1\right)^{11/2}}\,,\\[1ex]
E^z(6)&=\frac{6581763 \,e_t^{10}+53772430 \,e_t^8+70576832 \,e_t^6+18257424 \,e_t^4+587456 \,e_t^2-1280}{161280 \,\pi  \,e_t^2 \left(\,e_t^2-1\right)^8}\nl
&+\frac{\left(1575 \,e_t^{10}+58140 \,e_t^8+217360 \,e_t^6+169856 \,e_t^4+28032 \,e_t^2+512\right) \arccos\left(-\frac{1}{\,e_t}\right)}{512 \,\pi  \left(\,e_t^2-1\right)^{17/2}}\,.
\end{align}
\end{subequations}
Note that for $E^x(n)$ and $E^z(n)$, $n$ should be even, and odd for $E^y(n)$. Otherwise, this strategy does not work.

\section{List of Partial results}\label{sec:partialresult}
\begin{align}
\Delta\cE^{(l)}_{(0)}&=\frac{M\,\nu^2}{c^8\,j^{10}}\,\vast[\,\frac{3136 \,e_t^6}{45}+\frac{297 \,\pi ^3 \,e_t^5}{20}+\left(\frac{88576 \,\pi ^2}{675}-\frac{64}{45}\right) \,e_t^4\nl
&+\left(\frac{11741 \,\pi ^3}{24}-\frac{2755 \,\pi ^5}{64}\right) e_t^3+\left(-\frac{27776}{225}+\frac{844288 \,\pi ^2}{675}-\frac{280576 \,\pi ^4}{2625}\right) e_t^2\nl
&+\left(\frac{202289 \,\pi ^3}{96}-\frac{3419707 \,\pi ^5}{1152}+\frac{17885 \,\pi ^7}{64}\right) e_t+\left(\frac{55936}{1575}+\frac{421888 \,\pi ^2}{135}-\frac{509827072 \,\pi ^4}{212625}+\frac{104726528 \,\pi ^6}{496125}\right)\nl
&+\frac{1}{e_t}\Bigg(\frac{3553649 \,\pi ^3}{960}-\frac{154014913 \,\pi ^5}{4608}+\frac{102671023 \,\pi ^7}{3200}-\frac{47703411 \,\pi ^9}{16384}\Bigg)\nl
&+\frac{1}{e_t^2}\Bigg(\frac{1856}{189}+\frac{4238336 \,\pi ^2}{945}-\frac{1355776 \,\pi ^4}{81}+\frac{7286226944 \,\pi ^6}{893025}-\frac{719847424 \,\pi ^8}{1091475}\Bigg)\nl
&+\frac{1}{e_t^3}\Bigg(\frac{36730841 \,\pi ^3}{7680}-\frac{7602883699 \,\pi ^5}{46080}+\frac{24445251593 \,\pi ^7}{38400}-\frac{406754528303 \,\pi ^9}{819200}+\frac{2879946531 \,\pi ^{11}}{65536}\Bigg)\nl
&+\frac{1}{e_t^4}\Bigg(\frac{5312}{1485}+\frac{5176832 \,\pi ^2}{945}-\frac{905998336 \,\pi ^4}{14175}+\frac{12913147904 \,\pi ^6}{127575}-\frac{209391976448 \,\pi ^8}{5457375}+\frac{2021654528 \,\pi ^{10}}{693693}\Bigg)\nl
&+\frac{1}{e_t^5}\Bigg(\frac{86668951 \,\pi ^3}{15360}-\frac{191346755029 \,\pi ^5}{368640}+\frac{426736282133 \,\pi ^7}{76800}\nl
&\quad-\frac{440356873312201 \,\pi ^9}{29491200}+\frac{2142139802884279 \,\pi ^{11}}{206438400}-\frac{4738568225391 \,\pi ^{13}}{5242880}\Bigg)+\cO(1/e_t^6)\Bigg]\,,
\end{align}
\begin{align}
\Delta\cE^{(l)}_{(1)}&=\frac{M\,\nu^2}{c^{10}\,j^{12}}\,\vast\{e_t^8\,\Bigg(-\frac{288256}{315}+\frac{185824 \,\,\nu}{315}\Bigg)+e_t^7\,\Bigg(\frac{9216 \,\pi }{35}-\frac{110367 \,\pi^3}{560}+\frac{873 \,\pi^3 \,\nu }{7}\Bigg)\nl
&+e_t^6\,\Bigg[\frac{467596}{315}-\frac{6344704 \,\pi^2}{4725}+\frac{2898 \,\zeta
(3)}{5}+\left(\frac{81856}{315}+\frac{1715456 \,\pi^2}{1575}\right) \,\nu\Bigg]\nl
&+e_t^5\,\Bigg[\frac{2816 \,\pi }{225}-\frac{216981733 \,\pi^3}{36960}+\frac{514505 \,\pi ^5}{896}+\left(\frac{682223\,\pi^3}{168}-\frac{315785 \,\pi^5}{896}\right) \,\nu\Bigg]\nl
&+e_t^4\,\Bigg[\frac{3920593}{3150}-\frac{22037087 \,\pi ^2}{3780}+\frac{177987584 \,\pi ^4}{202125}+\frac{54936 \,\zeta(3)}{5}-\frac{8883 \,\pi ^2 \,\zeta (3)}{8}-\frac{118017 \,\zeta(5)}{8}\nl
&\quad+\left(-\frac{1670624}{1575}+\frac{49979392 \,\pi^2}{4725}-\frac{5280768 \,\pi^4}{6125}\right) \,\nu\Bigg]\nl
&+e_t^3\,\Bigg[-\frac{79616 \,\pi }{135} -\frac{202204565099 \,\pi^3}{11088000}+\frac{2230868679727 \,\pi^5}{57657600}-\frac{7622335 \,\pi^7}{2048}\nl
&\quad+\left(\frac{6238315 \,\pi^3}{336}-\frac{42527319 \,\pi^5}{1792}+\frac{2272095 \,\pi^7}{1024}\right) \,\nu\Bigg]\nl
&+e_t^2\,\Bigg[-\frac{5903108}{2205}+\frac{2185941973 \,\pi^2}{226800}+\frac{5354923998451 \,\pi^4}{523908000}-\frac{11762012782592 \,\pi^6}{8442559125}+\frac{76944 \,\zeta (3)}{5}\nl
&\quad-\frac{443177\,\pi^2 \,\zeta(3)}{8}+\frac{174195 \,\pi ^4 \,\zeta (3)}{64}-\frac{5887923 \,\zeta(5)}{8}+\frac{3857175 \,\pi^2\,\zeta (5)}{64}+\frac{47405925 \,\zeta(7)}{64}\nl
&\quad+\left(-\frac{1834496}{11025}+\frac{137875712 \,\pi^2}{4725}-\frac{28483790848 \,\pi^4}{1488375}+\frac{1145110528 \,\pi ^6}{694575}\right) \,\nu\Bigg]\nl
&+e_t\,\Bigg[\frac{12832 \,\pi }{45}-\frac{13673548523 \,\pi^3}{1056000}+\frac{2331631085671213 \,\pi^5}{5765760000}-\frac{181467504528103 \,\pi^7}{429977600}+\frac{1265629617 \,\pi^9}{32768}\nl
&\quad+\left(\frac{246285677 \,\pi^3}{6720}-\frac{17308193471 \,\pi^5}{64512}+\frac{12669225577 \,\pi^7}{51200}-\frac{732535083 \,\pi^9}{32768}\right)
\,\nu\Bigg]\nl
&+\Bigg[\frac{1423722529}{2116800}+\frac{735046449181 \,\pi^2}{19051200}-\frac{14545675750451 \,\pi^4}{41912640000}-\frac{1052053473232830449 \,\pi ^6}{38903312448000}\nl
&\quad+\frac{9956950540288 \,\pi^8}{2916520425}+\frac{10752 \,\zeta (3)}{5}-\frac{1682177 \,\pi ^2 \,\zeta (3)}{4}+\frac{39521979 \,\pi
^4 \,\zeta (3)}{160}-\frac{9183825 \,\pi ^6 \,\zeta (3)}{1024}\nl
&\quad-\frac{22348923 \,\zeta(5)}{4}+\frac{175025907 \,\pi^2 \,\zeta (5)}{32}-\frac{512457435 \,\pi^4 \,\zeta(5)}{2048}+\frac{2151124857 \,\zeta(7)}{32}-\frac{10497111975 \,\pi^2 \,\zeta(7)}{2048}\nl
&\quad-\frac{126709233525 \,\zeta(9)}{2048}+\Bigg(\frac{804128}{3675}+\frac{1533979136 \,\pi
^2}{33075}-\frac{202340205568 \,\pi ^4}{1488375}+\frac{26006683648 \,\pi
^6}{416745}\nl
&\quad-\frac{12671254528 \,\pi ^8}{2546775}\Bigg) \,\nu\Bigg]\nl
&+\frac{1}{e_t}\,\Bigg[\frac{8 \,\pi }{3}-\frac{13799463626119 \,\pi^3}{14370048000}+\frac{108826027744604749 \,\pi^5}{62270208000}-\frac{21125048445944211030463 \,\pi^7}{2628134885376000}\nl
&\quad+\frac{5848151327145833\,\pi ^9}{896860160}-\frac{152039294253 \,\pi^{11}}{262144}+\Bigg(\frac{341022193 \,\pi^3}{6720}-\frac{878042967227 \,\pi^5}{645120}\nl
&\quad+\frac{2979531597611 \,\pi^7}{614400}-\frac{42327561853999 \,\pi^9}{11468800}+\frac{21336464457 \,\pi ^{11}}{65536}\Bigg)
\,\nu\Bigg]\nl
&+\frac{1}{e_t^2}\,\Bigg[\frac{334356791}{3326400}+\frac{199332473447 \,\pi ^2}{3386880}-\frac{373142823899 \,\pi^4}{1587600}+\frac{3907498382463157 \,\pi^6}{424569600000}\nl
&\quad+\frac{584827839010454989447 \,\pi^8}{6570337213440000}-\frac{6152058407223296 \,\pi ^{10}}{541111756185}-\frac{27597339 \,\pi^2
\,\zeta (3)}{20}+\frac{1213492203 \,\pi^4 \,\zeta (3)}{320}\nl
&\quad-\frac{6498156357 \,\pi^6 \,\zeta(3)}{5120}+\frac{3215354373 \,\pi^8 \,\zeta (3)}{81920}-\frac{366650361 \,\zeta(5)}{20}+\frac{5374036899 \,\pi^2 \,\zeta(5)}{64}\nl
&\quad-\frac{1812985623603 \,\pi^4 \,\zeta(5)}{51200}+\frac{10171018935 \,\pi^6 \,\zeta(5)}{8192}+\frac{66048647049 \,\zeta(7)}{64}-\frac{7427392716051 \,\pi^2 \,\zeta(7)}{10240}\nl
&\quad+\frac{525021435477 \,\pi^4 \,\zeta(7)}{16384}-\frac{89655063257529 \,\zeta (9)}{10240}+\frac{10562439115305 \,\pi^2 \,\zeta(9)}{16384}\nl
&\quad+\frac{126935300600955\,\zeta(11)}{16384}+\Bigg(\frac{704192}{10395}+\frac{78030592\,\pi^2}{1323}-\frac{53212499968 \,\pi ^4}{99225}+\frac{137693298688 \,\pi^6}{178605}\nl
&\quad-\frac{10766761459712 \,\pi^8}{38201625}+\frac{308738523136 \,\pi^{10}}{14567553}\Bigg) \,\nu\Bigg]\nl
&+\frac{1}{e_t^3}\,\Bigg[\frac{1054 \,\pi }{135}+\frac{212624533804289 \,\pi^3}{36951552000}+\frac{11574755701298686367 \,\pi^5}{2490808320000}-\frac{25790664698470750454305469 \,\pi^7}{394220232806400000}\nl
&\quad+\frac{2025121269359422584900830537 \,\pi^9}{10638690016002048000}-\frac{5826127395266536829411 \,\pi^{11}}{42941664460800}+\frac{124362967035879 \,\pi^{13}}{10485760}\nl
&\quad+\Bigg(\frac{6592502233 \,\pi^3}{107520}-\frac{22857379501687 \,\pi^5}{5160960}+\frac{52153564678267 \,\pi^7}{1228800}-\frac{14962473720996481 \,\pi^9}{137625600}\nl
&\quad+\frac{1787772436877977 \,\pi^{11}}{24084480}-\frac{13523632850871 \,\pi^{13}}{2097152}\Bigg) \,\nu\Bigg]+\cO(1/e_t^4)\vast\}\,.
\end{align}
\begin{align}\label{deltaE23}
&\Delta\cE_{(2)}^{(l)}=\frac{M\,\nu^2\,(1-4\nu)}{c^{10}\,j^{12}}\Bigg[\frac{81856\, e_t^8}{945}+\frac{4053 \,\pi ^3 \,e_t^7}{160}+\left(\frac{50560}{189}+\frac{275456\,\,\pi ^2}{945}\right) e_t^6\nl
&+\left(\frac{28654757 \,\pi ^3}{20160}-\frac{1292765\,\pi ^5}{10752}\right) e_t^5+\left(-\frac{192064}{525}+\frac{4470784\,\pi ^2}{945}-\frac{20166656 \,\pi ^4}{55125}\right) e_t^4\nl
&+\left(\frac{163927481 \,\pi ^3}{16128}-\frac{795673307 \,\pi ^5}{64512}+\frac{2352595 \,\pi ^7}{2048}\right) e_t^3\nl
&+\left(-\frac{167936}{1225}+\frac{261379072\,\pi ^2}{14175}-\frac{53227319296 \,\pi ^4}{4465125}+\frac{10682630144\,\pi ^6}{10418625}\right) e_t^2\nl
&+\left(\frac{270990515\,\pi ^3}{10752}-\frac{16781634175 \,\pi ^5}{86016}+\frac{3732414369\,\pi ^7}{20480}-\frac{2161208385\,\pi ^9}{131072}\right) e_t\nl
&+\left(\frac{3086144}{33075}+\frac{471584768\,\pi ^2}{14175}-\frac{71593908224\,\pi ^4}{637875}+\frac{1672914731008 \,\pi ^6}{31255875}-\frac{4699193344 \,\pi ^8}{1091475}\right)\nl
&+\frac{1}{e_t}\Bigg(\frac{9587828293\,\pi ^3}{258048}-\frac{3207169211179\,\pi ^5}{2580480}+\frac{701408449909 \,\pi ^7}{147456}-\frac{509401541368309 \,\pi ^9}{137625600}\nl
&+\frac{171695021883\,\pi ^{11}}{524288}\Bigg)+\frac{1}{e_t^2}\Bigg(\frac{255104}{10395}+\frac{288994304 \,\pi ^2}{6615}-\frac{52566433792 \,\pi ^4}{99225}+\frac{758425714688 \,\pi ^6}{893025}\nl
&-\frac{12350325784576 \,\pi ^8}{38201625}+\frac{1074396135424 \,\pi ^{10}}{43702659}\Bigg)+\frac{1}{e_t^3}\Bigg(\frac{118105384883 \,\pi ^3}{2580480}-\frac{96013915997417 \,\pi ^5}{20643840}\nl
&+\frac{380628106184173 \,\pi ^7}{7372800}-\frac{232396062526713293 \,\pi ^9}{1651507200}+\frac{1136196879582428477 \,\pi ^{11}}{11560550400}\nl
&-\frac{359312576835069 \,\pi ^{13}}{41943040}\Bigg)+\cO(1/e_t^4)\Bigg]\,,\\[1.5ex]
&\Delta\cE_{(3)}^{(l)}=\frac{M\,\nu^2\,(1-4\nu)}{c^{10}\,j^{12}}\Bigg[\,\frac{512\, e_t^8}{135}+\frac{15 \,\,\pi^3 \,e_t^7}{16}+\left(\frac{1024}{135}+\frac{2048\,\pi^2}{225}\right) e_t^6+\left(\frac{10801 \,\pi ^3}{288}-\frac{1225 \,\pi ^5}{384}\right) e_t^5\nl
&+\left(-\frac{1024}{75}+\frac{14336 \,\pi ^2}{135}-\frac{65536 \,\pi ^4}{7875}\right) e_t^4+\left(\frac{226961 \,\pi ^3}{1152}-\frac{186277 \,\pi ^5}{768}+\frac{46305 \,\pi ^7}{2048}\right) e_t^3\nl
&+\left(-\frac{1024}{525}+\frac{641024 \,\pi ^2}{2025}-\frac{130777088 \,\pi ^4}{637875}+\frac{1048576 \,\pi ^6}{59535}\right) e_t^2\nl
&+\left(\frac{1521919 \,\pi ^3}{3840}-\frac{9146851 \,\pi ^5}{3072}+\frac{282281979 \,\pi ^7}{102400}-\frac{8164233 \,\pi ^9}{32768}\right) e_t\nl
&+\left(\frac{4096}{1575}+\frac{7026688 \,\pi ^2}{14175}-\frac{985563136 \,\pi ^4}{637875}+\frac{214433792 \,\pi ^6}{297675}-\frac{4194304 \,\pi ^8}{72765}\right)\nl
&+\frac{1}{e_t}(\frac{9909613 \,\pi ^3}{18432}-\frac{1446169151 \,\pi ^5}{92160}+\frac{42478149259 \,\pi ^7}{737280}-\frac{217594014011 \,\pi ^9}{4915200}+\frac{1025047023 \,\pi ^{11}}{262144})\nl
&+\frac{1}{e_t^2}\Bigg(\frac{1024}{1485}+\frac{83968 \,\pi ^2}{135}-\frac{2523136 \,\pi ^4}{405}+\frac{1194852352 \,\pi ^6}{127575}-\frac{2716860416 \,\pi ^8}{779625}+\frac{33554432 \,\pi ^{10}}{127413}\Bigg)\nl
&+\frac{1}{e_t^3}\Bigg(\frac{118881539 \,\pi ^3}{184320}-\frac{38059061669 \,\pi ^5}{737280}+\frac{3865364478731 \,\pi ^7}{7372800}\nl
&\quad-\frac{27120739832417 \,\pi ^9}{19660800}+\frac{10474810693061 \,\pi ^{11}}{11010048}-\frac{173590231815 \,\pi ^{13}}{2097152}\Bigg)+\cO(1/e_t^4)\Bigg]\,,
\end{align}
\begin{align}\label{deltaE4}
\Delta\cE_{(4)}&=\frac{M\,\nu^2}{c^{11}\,j^{13}}\Bigg[\left(\frac{163787 \,e_t^6}{450}+\frac{180379 \,e_t^4}{75}+\frac{166996 \,e_t^2}{75}+\frac{62744}{225}\right) \sqrt{e_t^2-1}\notag\\[1ex]
&+\left(\frac{297\, e_t^8}{10}+\frac{14008\, e_t^6}{15}+\frac{41368 \,e_t^4}{15}+\frac{4352\, e_t^2}{3}+\frac{512}{5}\right) \arccos\left(-\frac{1}{e_t}\right)\Bigg]\notag\\[1ex]
&\times\Bigg[\frac{2\,\pi^2}{3}-\frac{214}{105}\,\left(\,\log \left(\frac{2\, n \,r_0}{G\,M\,c\,e_t}\right)+\gamma\right)-\frac{116761}{29400}\Bigg]\,,
\end{align}
\begin{align}
&\Delta\cE^{(l)}_{(5)}=\frac{M\,\nu^2}{c^{11}\,j^{13}}\,\Bigg[\,e_t^8 \left(\frac{10593 \,\gamma  \,\pi }{350}-107 \,\pi +\frac{31779}{350} \,\pi  \,\log (2)\right)+\frac{54784 \,e_t^7}{354375} \Bigg(5190 \,\gamma -5047-5190 \,\log (2)\Bigg)\nl
&+\frac{107 \,e_t^6 }{151200}\Bigg(-2603475 \,\pi  \,\zeta (3)+1344768 \,\gamma  \,\pi -2290582 \,\pi +4034304 \,\pi  \,\log (2)\Bigg)\nl
&+\frac{54784 \,e_t^5}{2480625} \Bigg(-443880 \,\zeta (3)+291795 \,\gamma +45734-291795 \,\log (2)\Bigg)\nl
&+\frac{107 \,e_t^4}{302400} \Bigg(-351258810 \,\pi  \,\zeta (3)+374243625 \,\pi  \,\zeta (5)+7942656 \,\gamma  \,\pi -6335000 \,\pi +23827968 \,\pi  \,\log (2)\Bigg)\nl
&+\frac{6848 \,e_t^3}{7441875} \Bigg(-223001760 \,\zeta (3)+220907520 \,\zeta (5)+8625330 \,\gamma +8763361-8625330 \,\log (2)\Bigg)\nl
&+\frac{107 \,e_t^2 }{193536000}\Bigg(-2243993740800 \,\pi  \,\zeta (3)+27140132920320 \,\pi  \,\zeta (5)-25257903843375 \,\pi  \,\zeta (7)\nl
&\quad\quad+2673868800 \,\gamma  \,\pi +186045664 \,\pi +8021606400 \,\pi  \,\log (2)\Bigg)\nl
&+\frac{6848 \,e_t}{573024375} \Bigg(-101159732520 \,\zeta (3)+631953315840 \,\zeta (5)-531449856000 \,\zeta (7)\nl
&\quad\quad +140502285 \,\gamma +277115017-140502285 \,\log (2)\Bigg)\nl
&+\Bigg(-\frac{17869266109 \,\pi  \,\zeta (3)}{3600}+\frac{1338717347453 \,\pi  \,\zeta (5)}{4800}-\frac{483420808341489 \,\pi  \,\zeta (7)}{204800}+\frac{137148534994779 \,\pi  \,\zeta (9)}{65536}\nl
&\quad\quad+\frac{54784 \,\gamma  \,\pi }{525}+\frac{2277666521 \,\pi }{16128000}+\frac{54784}{175} \,\pi  \,\log (2)\Bigg)\nl
&+\frac{428}{7449316875 \,e_t} \Bigg(-63394926154560 \,\zeta (3)+1492742225694720 \,\zeta (5)-6265138721587200 \,\zeta (7)\nl
&\quad\quad+4835860807680000 \,\zeta (9)+401531130 \,\gamma +1718774371-401531130 \,\log (2)\Bigg)\nl
&-\frac{15301}{3538944000 \,e_t^2} \Bigg(2814246420480 \,\pi  \,\zeta (3)-505641831137280 \,\pi  \,\zeta (5)+15738675691455360 \,\pi  \,\zeta (7)\nl
&\quad\quad-113566524198701760 \,\pi  \,\zeta (9)+98396833403743875 \,\pi  \,\zeta (11)+52736 \,\pi \Bigg)\nl
&+\frac{428}{4469590125 \,e_t^3} \Bigg(-80247248257176 \,\zeta (3)+5268405431697408 \,\zeta (5)-66851431345520640 \,\zeta (7)\nl
&\quad\quad+230673046606184448 \,\zeta (9)-169009790268211200 \,\zeta (11)+2171169 \,\gamma +1653152-2171169 \,\log (2)\Bigg)\nl
&+\cO(1/e_t^4)\Bigg]\,,
\end{align}
where $\zeta(z)$ is Riemann-Zeta function.
\section{Total Energy radiation}\label{sec:fullresult}
In this Appendix, we collect all contributions of total energy radiation $\Delta\cE=\Delta\cE_\text{inst}+\Delta\cE_\text{hered}$  upto 3PN and $1/j^{15}$ order including both instantaneous and hereditary contributions. Note that, to be consistent, we expand the contributions of which exact form are known (instantaneous part and $\Delta E_{(4)}$) and discard $1/j^{16}$ contributions of $\Delta E^{(l)}_{(2)}$ and $\Delta E^{(l)}_{(3)}$ already computed in Eq.~(\ref{deltaE23}). Because of the size of the expression, we introduce shorthand notations,
\begin{align}
p:=\sqrt{\left(1+E+\frac{1}{2}\,\nu\,E^2\right)^2-1}\,,
\end{align}
($p$ is usually called the linear momentum at infinity) and $J$ (do not be confused with the physical angular momentum $\mathcal{J}$) defined as
\begin{align}
\frac{1}{J^n}:=\frac{M\,\nu^2}{c^{n-2}}\,\frac{p^{7-n}}{j^n}\,,
\end{align}
so that whatever value of the positive integer $n$ is, $\frac{1}{J^n}$ has a dimension of energy $\mathcal{E}\sim \text{mass}\, \frac{\text{length}^2}{\text{time}^2}$, and is formally as small as $\cO(\frac{1}{c^5})$ (\textit{i.e.} 2.5PN order, the leading order of energy radiation) and $\cO(G^n)$, while $p$ is dimensionless and as small as $\cO(\frac{1}{c})$ without entailing $G$, hence which is going to serve as a PN parameter. Every coefficients of $1/J^n$ are dimensionless and polynomials of $p$, of which the highest power is 6 (that is, 3PN).
\begin{align}
&\Delta\cE=\frac{\,\pi}{J^3}\,\Bigg[\frac{37}{15}+p^2 \left(\frac{1357}{840}-\frac{74 \nu}{15}\right)+p^4\left(\frac{27953}{10080}-\frac{839 \nu}{420}+\frac{37 \nu^2}{5}\right)+p^6 \left(-\frac{676273}{354816}-\frac{2699 \nu}{504}+\frac{321 \nu^2}{280}-\frac{148 \nu^3}{15}\right)\Bigg]\nl
&+\frac{1}{J^4}\,\Bigg[\frac{1568}{45}+p^2 \left(\frac{18608}{525}-\frac{1424 \nu}{15}\right)+p^3\frac{3136 }{45}+p^4 \left(\frac{220348}{11025}-\frac{31036 \nu}{525}+172 \nu^2\right)\nl
&\quad+p^5 \left(\frac{1216}{105}-\frac{2848 \nu}{15}\right)+p^6 \left(-\frac{151854}{13475}-\frac{1223594 \nu}{33075}+\frac{164 \nu^2}{3}-\frac{2366 \nu^3}{9}\right)\Bigg]\nl
&+\frac{\,\pi}{J^5}\,\Bigg[\frac{122}{5}+p^2 \left(\frac{13831}{280}-\frac{933 \nu}{10}\right)+p^3 \frac{297 \,\pi ^2}{20}+p^4 \left(-\frac{64579}{5040}-\frac{187559 \nu}{1680}+\frac{2067 \nu^2}{10}\right)+p^5 \left(\frac{9216}{35}-\frac{24993 \pi ^2}{1120}-\frac{15291\pi^2 \nu}{280} \right)\nl
&\quad+p^6 \left(\frac{29573617463}{310464000}-\frac{10593 \log(\cfrac{p}{2})}{350}+\frac{99 \pi ^2}{10}+\frac{76897 \nu}{480}-\frac{4059}{640} \pi ^2 \nu+\frac{12269 \nu^2}{80}-\frac{1823 \nu^3}{5}\right)\Bigg]\nl
&+\frac{1}{J^6}\,\Bigg[\frac{4672}{45}+p^2 \left(\frac{142112}{315}-\frac{26464 \nu}{45}\right)+p^3 \left(\frac{9344}{45}+\frac{88576 \pi ^2}{675}\right)+p^4 \left(-\frac{293992}{1701}-\frac{6732728 \nu}{4725}+\frac{24424 \nu^2}{15}\right)\nl
&\quad+p^5 \left(\frac{56708}{105}+\frac{1024 \pi ^2}{135}+\frac{2898 \zeta (3)}{5}-\frac{52928 \nu}{45}-\frac{3014912 \pi ^2 \nu}{4725}\right)\nl
&\quad+p^6 \left(\frac{36589282372}{11694375}-\frac{18955264 \log(2p)}{23625}+\frac{177152 \pi ^2}{675}+\frac{875976284 \nu}{297675}-\frac{212216 \pi ^2 \nu}{1575}+\frac{4201976 \nu^2}{1575}-\frac{150892 \nu^3}{45}\right)\Bigg]\nl
&+\frac{\,\pi}{J^7}\,\Bigg[\frac{85}{3}+p^2 \left(\frac{2259}{8}-265 \,\nu\right)+p^3 \left(\frac{1579 \,\pi ^2}{3}-\frac{2755 \,\pi ^4}{64}\right)+p^4 \left(\frac{19319}{378}-\frac{432805 \nu}{336}+\frac{7605 \nu^2}{8}\right)\nl
&\quad+p^5 \left(\frac{210176}{225}-\frac{13138915 \pi ^2}{7392}+\frac{689985 \pi ^4}{3584}-\frac{23514}{7} \pi ^2 \nu+\frac{30285}{112} \pi ^4 \nu\right)\nl
&\quad+p^6 \left(\frac{37546579757}{8467200}-\frac{337906 \log(\cfrac{p}{2})}{315}+\frac{3158 \pi ^2}{9}-\frac{58957 \zeta
(3)}{32}+\frac{68898691 \nu}{36288}-\frac{51947}{384} \pi ^2 \nu+\frac{1419153 \nu^2}{448}-\frac{13955
\nu^3}{6}\right)\Bigg]\nl
&+\frac{1}{J^8}\,\Bigg[\frac{3104}{75}+p^2 \left(\frac{625808}{525}-\frac{21136 \nu}{25}\right)+p^3 \left(\frac{6208}{75}+\frac{68096 \pi ^2}{45}-\frac{280576
\pi ^4}{2625}\right)+p^4 \left(\frac{29175292}{14175}-\frac{1861268 \nu}{225}+\frac{104404 \nu^2}{25}\right)\nl
&\quad+p^5 \Bigg(\frac{495031}{210}+\frac{3097739 \pi^2}{1260}+\frac{1062912 \pi ^4}{13475}+12726
\zeta (3)-\frac{8883 \pi ^2 \zeta
(3)}{8}-\frac{118017 \zeta (5)}{8}\nl
&\quad-\frac{42272
\nu}{25}-\frac{6647552}{525} \pi ^2 \nu+\frac{15586304 \pi ^4 \nu}{18375}\Bigg)\nl
&\quad+p^6 \Bigg(\frac{46042764886}{1091475}-\frac{2081792
\log(2p)}{225}+\frac{136192 \pi^2}{45}-\frac{60043264 \zeta
(3)}{6125}\nl
&\quad-\frac{8449234 \nu
}{4725}-\frac{19516}{25} \pi ^2 \nu+\frac{14002412 \nu^2}{525}-\frac{934166
\nu^3}{75}\Bigg)\Bigg]\nl
&+\frac{\,\pi}{J^9}\,\Bigg[p^2 \left(\frac{13447}{40}-\frac{1127 \,\nu}{6}\right)+p^3 \left(\frac{17213 \,\pi ^2}{6}-\frac{873523 \,\pi ^4}{288}+\frac{17885 \,\pi ^6}{64}\right)+p^4 \left(\frac{11947909}{6480}-\frac{2838577 \,\nu}{720}+\frac{5733 \,\nu^2}{4}\right)\nl
&\quad+p^5 \left(\frac{80128}{135}-\frac{1554265673 \pi ^2}{126000}+\frac{346721827097 \pi ^4}{16473600}-\frac{4078795 \pi
   ^6}{2048}-\frac{126707}{4} \pi ^2 \nu+\frac{34662173 \pi ^4 \nu}{1152}-\frac{2811865 \pi ^6 \nu}{1024}\right)\nl
&\quad+p^6 \Bigg(\frac{111152851823}{3628800}-\frac{263113 \log(\cfrac{p}{2})}{45}+\frac{17213 \pi ^2}{9}-\frac{93466961 \zeta
   (3)}{720}+\frac{8474935 \zeta (5)}{64}\nl
&\quad-\frac{343378331 \nu}{25920}-\frac{96145}{384} \pi ^2 \nu+\frac{3277505 \nu^2}{192}-\frac{86681 \nu^3}{16}\Bigg)\Bigg]\nl
&+\frac{1}{J^{10}}\,\Bigg[-\frac{15488}{1575}+p^2 \left(\frac{1897024}{3675}-\frac{123328 \nu}{525}\right)+p^3 \left(-\frac{30976}{1575}+\frac{3042304 \pi ^2}{675}-\frac{76079104 \pi ^4}{30375}+\frac{104726528 \pi ^6}{496125}\right)\nl
&\quad+p^4 \left(\frac{916671536}{99225}-\frac{148687088 \nu}{11025}+\frac{644624 \nu^2}{175}\right)\nl
&\quad+p^5 \Bigg(\frac{1297879}{1225}+\frac{852398347 \pi ^2}{32400}-\frac{193061153711 \pi ^4}{174636000}-\frac{985608224768 \pi^6}{2814186375}+39102 \zeta (3)-\frac{460943 \pi ^2 \zeta (3)}{8}+\frac{174195 \pi ^4 \zeta (3)}{64}\nl   
&\quad-\frac{6123957
\zeta (5)}{8}+\frac{3857175 \pi ^2 \zeta (5)}{64}+\frac{47405925 \zeta (7)}{64}-\frac{246656 \nu}{525}-\frac{35044864}{525} \pi ^2 \nu+\frac{2186350592 \pi ^4 \nu}{70875}-\frac{417267712 \pi ^6 \nu}{165375}\Bigg)\nl
&\quad+p^6 \Bigg(\frac{14145635478904}{81860625}-\frac{651053056 \log(2p)}{23625}+\frac{6084608 \pi ^2}{675}-\frac{16280928256
\zeta (3)}{70875}+\frac{11205738496 \zeta (5)}{55125}\nl
&\quad-\frac{36786359432 \nu}{297675}-\frac{9184}{225} \pi ^2\nu+\frac{304241296 \nu^2}{3675}-\frac{29597608 \nu^3}{1575}\Bigg)\Bigg]\nl
&+\frac{\,\pi}{J^{11}}\,\Bigg[p^3 \left(\frac{24717 \,\pi ^2}{5}-\frac{2235121 \,\pi ^4}{64}+\frac{25779537 \,\pi ^6}{800}-\frac{47703411 \,\pi ^8}{16384}\right)+p^4 \left(\frac{5839651}{2016}-\frac{258051 \,\nu}{80}+\frac{5481 \,\nu^2}{8}\right)\nl
&\quad+p^5 \Bigg(-\frac{41509919 \pi ^2}{2800}+\frac{14329698290513 \pi ^4}{45760000}-\frac{13115655312543 \pi^6}{42997760}+\frac{3631907727 \pi ^8}{131072}-\frac{1012779}{10} \pi ^2 \nu\nl
&\quad+\frac{34063737}{64} \pi ^4 \nu
-\frac{11886663789 \pi ^6 \nu}{25600}+\frac{341480853 \pi ^8 \nu}{8192}\Bigg)\nl
&\quad+p^6 \Bigg(\frac{8986587257}{115200}-\frac{251878 \log(\cfrac{p}{2})}{25}+\frac{16478 \pi^2}{5}-\frac{239157947 \zeta
(3)}{160}+\frac{12215817747 \zeta (5)}{800}-\frac{114395585661 \zeta (7)}{8192}\nl
&\quad-\frac{465343901 \nu}{6720}+\frac{255717}{640} \pi ^2 \nu+\frac{10035909 \nu^2}{320}-\frac{42399 \nu^3}{8}\Bigg)\Bigg]\nl
&+\frac{1}{J^{12}}\,\Bigg[\frac{928}{189}+p^2 \left(-\frac{911024}{6615}+\frac{8464 \nu}{189}\right)+p^3 \left(\frac{1856}{189}+\frac{4238336 \pi ^2}{945}-\frac{1355776 \pi ^4}{81}+\frac{7286226944 \pi ^6}{893025}-\frac{719847424
\pi ^8}{1091475}\right)\nl
&\quad+p^4 \left(\frac{286020724}{59535}-\frac{5744404 \nu}{1323}+\frac{46468 \nu^2}{63}\right)\nl
&\quad+p^5 \Bigg(-\frac{121500019}{423360}+\frac{307739224253 \pi ^2}{3810240}-\frac{8813724626501 \pi
^4}{181440000}-\frac{2647984678792493 \pi ^6}{457686028800}+\frac{54187719655424 \pi ^8}{32081724675}\nl
&\quad+29106 \zeta
(3)-\frac{1908207 \pi ^2 \zeta (3)}{4}+\frac{79914933 \pi ^4 \zeta (3)}{320}-\frac{9183825 \pi ^6 \zeta
(3)}{1024}-\frac{25351893 \zeta (5)}{4}+\frac{353908989 \pi ^2 \zeta (5)}{64}\nl
&\quad-\frac{512457435 \pi ^4 \zeta
   (5)}{2048}+\frac{4349655639 \zeta (7)}{64}-\frac{10497111975 \pi ^2 \zeta (7)}{2048}-\frac{126709233525 \zeta
   (9)}{2048}+\frac{16928 \nu}{189}\nl
&\quad-\frac{98811392}{735} \pi ^2 \nu+\frac{2685110272 \pi ^4 \nu}{8505}-\frac{125715644416 \pi ^6 \nu}{893025}+\frac{7749763072 \pi ^8 \nu}{694575}\Bigg)\nl
&\quad+p^6 \Bigg(\frac{10229486279182}{38201625}-\frac{907003904 \log(2p)}{33075}+\frac{8476672 \pi ^2}{945}-\frac{290136064 \zeta
(3)}{189}+\frac{779626283008 \zeta (5)}{99225}\nl
&\quad-\frac{154047348736 \zeta (7)}{24255}-\frac{2203483862 \nu}{8505}+\frac{696344}{315} \pi ^2 \nu+\frac{27943012 \nu^2}{315}-\frac{2177662 \nu^3}{189}\Bigg)\Bigg]\nl
&+\frac{\,\pi}{J^{13}}\,\Bigg[p^3 \left(\frac{10549 \,\pi ^2}{4}-\frac{35497847 \,\pi ^4}{240}+\frac{297848551 \,\pi ^6}{480}-\frac{101390485757 \,\pi ^8}{204800}+\frac{2879946531 \,\pi ^{10}}{65536}\right)\nl
&\quad+p^5 \Bigg(\frac{2089356493 \pi ^2}{116640}+\frac{9128301066877073 \pi ^4}{5987520000}-\frac{30683901515370260411 \pi^6}{4322590272000}+\frac{51653285838124707 \pi^8}{8968601600}-\frac{268570906065 \pi^{10}}{524288}\nl
&\quad-\frac{5166909}{40} \pi ^2 \nu+\frac{2218568099}{640} \pi ^4 \nu-\frac{81672407509 \pi ^6 \nu}{6400}+\frac{56346284097807 \pi ^8 \nu}{5734400}-\frac{113792507829 \pi ^{10} \nu}{131072}\Bigg)\nl
&\quad+p^6 \Bigg(\frac{72363445187}{1075200}-\frac{161249 \log(\cfrac{p}{2})}{30}+\frac{10549 \pi ^2}{6}-\frac{3798269629 \zeta(3)}{600}+\frac{141137663381 \zeta(5)}{480}\nl
&\quad-\frac{243140348991507 \zeta (7)}{102400}+\frac{137148534994779 \zeta(9)}{65536}-\frac{3972009943 \nu}{60480}+\frac{208813\pi^2 \nu}{320} +\frac{1157409 \nu^2}{64}-\frac{29645 \nu^3}{16}\Bigg)\Bigg]\nl
&+\frac{1}{J^{14}}\,\Bigg[-\frac{10816}{3465}+p^2 \left(\frac{1338784}{17325}-\frac{4448 \nu}{231}\right)\nl
&\quad+p^3 \left(-\frac{21632}{3465}+\frac{312832 \pi^2}{315}-\frac{222912512 \pi ^4}{4725}+\frac{27701936128 \pi^6}{297675}-\frac{6236143616 \pi^8}{165375}+\frac{2021654528 \pi^{10}}{693693}\right)\nl
&\quad+p^4 \left(-\frac{1571460808}{1091475}+\frac{346442552 \nu}{363825}-\frac{127592 \nu^2}{1155}\right)\nl
&\quad+p^5 \Bigg(\frac{173028797}{1108800}+\frac{97371732173 \pi^2}{1128960}-\frac{12705804583 \pi^4}{35280}+\frac{106040792865406033
\pi^6}{990662400000}+\frac{5290677461059608407 \pi^8}{81115274240000}\nl
&\quad-\frac{5322592932069376 \pi
^{10}}{541111756185}-\frac{27597339 \pi^2 \zeta (3)}{20}+\frac{1213492203 \pi^4 \zeta (3)}{320}-\frac{6498156357
\pi^6 \zeta (3)}{5120}+\frac{3215354373 \pi^8 \zeta (3)}{81920}\nl
&\quad-\frac{366650361 \zeta (5)}{20}+\frac{5374036899
\pi ^2 \zeta (5)}{64}-\frac{1812985623603 \pi ^4 \zeta (5)}{51200}+\frac{10171018935 \pi ^6 \zeta
(5)}{8192}+\frac{66048647049 \zeta (7)}{64}\nl
&\quad-\frac{7427392716051 \pi ^2 \zeta (7)}{10240}+\frac{525021435477 \pi ^4
\zeta (7)}{16384}-\frac{89655063257529 \zeta (9)}{10240}+\frac{10562439115305 \pi^2 \zeta(9)}{16384}\nl
&\quad+\frac{126935300600955 \zeta (11)}{16384}-\frac{8896 \nu}{231}-\frac{25802496}{245} \pi ^2 \nu+\frac{15275511808 \pi ^4 \nu}{11025}-\frac{32249151488 \pi ^6 \nu}{14175}\nl
&\quad+\frac{11118153433088 \pi ^8
\nu}{12733875}-\frac{969316237312 \pi ^{10} \nu}{14567553}
\Bigg)\nl
&\quad+p^6 \Bigg(\frac{551319893540236}{5462832375}-\frac{66946048 \log(2p)}{11025}+\frac{625664 \pi ^2}{315}-\frac{47703277568
\zeta (3)}{11025}+\frac{2964107165696 \zeta (5)}{33075}\nl
&\quad-\frac{1334534733824 \zeta (7)}{3675}+\frac{1946853310464
\zeta (9)}{7007}-\frac{33565371524 \nu}{363825}+984 \pi ^2 \nu+\frac{103067368 \nu^2}{4851}-\frac{6077972 \nu^3}{3465}\Bigg)\Bigg]\nl
&+\frac{\,\pi}{J^{15}}\,\Bigg[p^3 \left(-\frac{272920219 \pi ^4}{960}+\frac{11072684623 \pi ^6}{2400}-\frac{52303041073283 \pi
   ^8}{3686400}+\frac{266066506940663 \pi ^{10}}{25804800}-\frac{4738568225391 \pi ^{12}}{5242880}\right)\nl
&\quad+p^5 \Bigg(\frac{5244099463 \pi ^2}{209952}+\frac{107378064314399 \pi ^4}{36288000}-\frac{1137117325791451946827 \pi^6}{19070251200000}+\frac{43996398056106875941927697 \pi^8}{231275869913088000}\nl
&\quad-\frac{5992795752733970871587 \pi^{10}}{42941664460800}+\frac{513752944703427 \pi^{12}}{41943040}-\frac{113399}{2} \pi ^2 \nu+\frac{35658158393
\pi ^4 \nu}{3360}-\frac{2572090864343 \pi ^6 \nu}{19200}\nl
&\quad+\frac{9846906270402487 \pi ^8 \nu}{25804800}-\frac{195459851677530583 \pi ^{10} \nu}{722534400}+\frac{7743141717597 \pi ^{12}\nu}{327680}\Bigg)\nl
&\quad+p^6 \Bigg(\frac{13990148029}{6912000}-\frac{29202463433\zeta(3)}{2400}+\frac{5246870699213 \zeta(5)}{2400}-\frac{41808589714402511 \zeta(7)}{614400}\nl
&\quad+\frac{201120530875501809 \zeta
(9)}{409600}-\frac{446094799380943713 \zeta (11)}{1048576}\Bigg)\Bigg]+\cO(1/c^{12}, G^{16})\,.
\end{align}
\end{widetext}

\nocite{*}
\bibliography{mybib}
\end{document}